\documentclass[twocolumn]{mn2e}
\usepackage{epsfig}

\title[The local star formation rate and radio luminosity density]
{The local star formation rate and radio luminosity density}
\author[Stephen Serjeant, Carlotta Gruppioni, Seb Oliver]{
Stephen Serjeant$^{1,4}$, Carlotta Gruppioni$^{2,4}$, 
Seb Oliver$^{3,4}$\vspace{0.3cm}\\
$^1$Unit for Space Sciences and Astrophysics, School of Physical
Sciences, University of Kent, Canterbury, Kent, CT2 7NR\\
$^2$Osservatorio Astronomico di Bologna, via Ranzani 1, 40127 Bologna, 
Italy\\
$^3$Astronomy Centre, CPES, University of Sussex, Falmer, Brighton BN1 9QJ\\
$^4$Astrophysics Group, Blackett Laboratory, Imperial College of 
Science Technology \& Medicine,Prince Consort
Rd.,London SW7 2BW
}
\date{}
 
\pagerange{\pageref{firstpage}--\pageref{lastpage}}
\pubyear{2001}
\volume{}

\begin{document}
\input ./BoxedEPS.tex
\SetEPSFDirectory{./}
\SetRokickiEPSFSpecial
\HideDisplacementBoxes

\label{firstpage}

\maketitle


\begin{abstract}
We present a new determination of the local volume-averaged 
star formation rate from the $1.4$ GHz
luminosity function of star forming galaxies. Our sample, taken
from the $B\le12$ 
Revised Shapley-Ames catalogue ($231$ normal spiral galaxies over 
effective area $7.1$ sr) has $\simeq100\%$ complete  
radio detections and is insensitive to dust
obscuration and cirrus 
contamination.
After removal of known active galaxies, the best-fit
Schechter function has a faint-end slope of $-1.27\pm0.07$ in 
agreement with the local H$\alpha$ luminosity function, 
characteristic luminosity $L_*=(2.6\pm0.7)\times10^{22}$ W Hz$^{-1}$
and density $\phi_*=(4.8\pm1.1)\times10^{-4}$ Mpc$^{-3}$.
The inferred local radio luminosity density of
$(1.73\pm0.37\pm0.03)\times10^{19}$ W Hz$^{-1}$ 
Mpc$^{-3}$ (Poisson noise, large scale structure fluctuations) 
implies a volume averaged star formation rate $\sim2\times$
larger than the Gallego et al. H$\alpha$ estimate, i.e. 
$\rho_{1.4{\rm GHz}}=(2.10\pm0.45\pm0.04)\times10^{-2}$ M$_\odot$
yr$^{-1}$ Mpc$^{-3}$ for
a Salpeter initial mass function from $0.1-125$ M$_\odot$ and Hubble
constant of $50$ km s$^{-1}$ Mpc$^{-1}$. 
We demonstrate that the Balmer decrement is a highly unreliable
extinction estimator, and argue that optical-UV SFRs are easily
underestimated, particularly at high redshift. 
\end{abstract}

\maketitle

\section{Introduction}\label{sec:introduction}
Some of the most ambitious and widely-cited extragalactic observations 
in recent years have been the constraints made on the star formation
history of the Universe. 
Madau et al. (1996) used the ultraviolet 
luminosity density as a tracer of the comoving star formation rate
(SFR), and by combining
$z\stackrel{>}{_\sim}3$ Lyman dropout surveys (Steidel et al. 1995)
with 
the $0.1<z<1$ Canada-France Redshift Survey
(CFRS; Lilly et al. 1996) they inferred a $z\sim1.5$ peak and 
subsequent redshift cut-off in the SFR. 
If correct this is a major 
result, implying an empirical determination of the dominant epoch of
metal production in the Universe 
(also suggestively close to the peak in QSO number density)
which can be compared directly with
hierarchical models of galaxy formation (Baugh et al. 1997).
The dereddened CFRS ultraviolet SFR quoted in Lilly et al. (1996)
appears to asymptote to the local 
(dereddened) H$\alpha$ estimate from Gallego et al. (1995). 
However ultraviolet estimates are very
sensitive to dust obscuration, with reddening corrections ranging from
factors of order $2$ (Madau et al. 1996) to $10$ (Meurer et
al. 1999). Indeed there have been several
claims that less reddening-dependent measures yield substantially
higher star formation rates: for example, sub-mm galaxy surveys both
behind lensing clusters (e.g. Blain et al 1999a,b; Ivison et al. 2000) 
and in the field (e.g. Hughes
et al 1998, Barger 
et al. 1999a,b, 
Lilly et al. 1999a,b, Eales et al. 1999, 2000, Peacock et al. 2000, 
Serjeant et al. 2001; Fox et
al. 2001 in preparation, Scott 
et al. 2001 in preparation) 
find star formation rates at $z>3$ comparable to
dereddened UV-selected samples, but the galaxy populations which
appear to comprise the sub-mm point sources overlap
little with the UV-selected star forming population; ISO observations
of the Hubble Deep Fields at $7\mu$m and $15\mu$m
(Rowan-Robinson et al. 1996, Oliver et al 2001 in preparation) find
systematically higher star formation rates than inferred at
$z\sim0.5-1$ in the optical; the H$\alpha$ luminosity density in the
CFRS (Tresse \& Maddox 1998) implies a $z\sim0.3$
star formation rate $2-3$ times
higher than that inferred from de-reddened ultraviolet estimates; and
the evolving radio luminosity density is also consistent with
comparable amounts of star formation from UV samples and inaccessible
to UV samples (Haarsma, Partridge, Windhorst \&
Richards 2000). 

In this paper we will show that the local H$\alpha$ star formation rate
is itself also a significant underestimate, by comparing it with the
local radio luminosity density of star forming galaxies. 
This paper is structured as follows. In section \ref{sec:method}
we present the sample used in this paper, and corresponding selection
criteria. Section \ref{sec:results} presents a calculation of the
local radio luminosity function and luminosity density.
Implications of these
results are discussed in section \ref{sec:discussion}.
We assume a Hubble constant of $H_0=50$ km s$^{-1}$
Mpc$^{-1}$,  $q_0=0.5$. 

\begin{figure}
\centering
  \ForceWidth{6in}
\vspace*{-1cm}
  \hSlide{-3.4cm}
  \BoxedEPSF{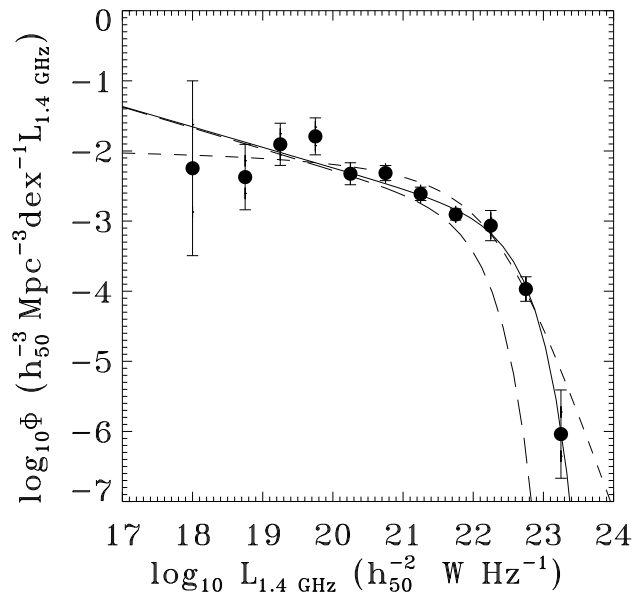}
\caption{\label{fig:rlf}
Local $1.4$ GHz radio luminosity function of Revised
Shapley-Ames spiral galaxies, with known AGN from V\'{e}ron-Cetty \&
V\'{e}ron (1993) catalogue excluded. Errors are the
$\Sigma1/V_{\rm max}\pm\sqrt{\Sigma1/V_{\rm max}^2}$ range in the
case of $>1$ galaxy in 
the bin; for bins with a single galaxy the errors represent
$\pm1\sigma$ limits on the appropriate Poisson distribution mean. 
The full line shows the Schechter
function fit described in the text, and the short dashed line shows the
original Condon (1989) luminosity function.
The long dashed line shows the best-fit H$\alpha$ luminosity function
of Gallego et al. (1995), scaled using equations \ref{eqn:radio2sfr}
and \ref{eqn:halpha2sfr}.}
\end{figure}

\section{Sample definition}\label{sec:method}
Unlike the far-IR, H$\alpha$ and UV, the radio luminosity density
traces both obscured and unobscured star formation, as well as being
free of cirrus contamination which affects far-IR estimates
at faint luminosities (Condon, Anderson \& Helou 1991).
At low radio frequencies
($\sim1$GHz) the luminosity is dominated by non-thermal emission,
plausibly from cosmic ray electrons accelerated by supernovae, which
can be calibrated to the recent star formation rate using the
supernova rate in the Galaxy (Condon 1992; section \ref{sec:rlf}). 
The low frequency radio luminosity density is therefore an ideal
estimator of the 
local volume-averaged star formation rate. 

The $B\leq12$ spiral galaxies in the Revised Shapley-Ames catalogue
were mapped at $1.4$GHz by Condon (1987) with the VLA to limiting flux
densities of (typically) $0.1-0.5$ mJy, obtaining detections of virtually
$100\%$ of the sample.
The survey spans a sufficiently large
cosmological volume for large scale structure density fluctuations to
be negligible in comparison with the shot noise on the luminosity
density and overall number density (Condon 1989; 
Oliver, Gruppioni \& Serjeant 1998):
using the Peacock \& Dodds (1994) power spectrum we estimate density
fluctuations of less than $2\%$ on the scale of the survey, which
agrees with the uniformity of the $<V/V_{\rm max}>$ statistic derived
by Condon (1989). Nevertheless, as a 
precaution the Local Group is excluded by eliminating the $10$
galaxies within $1.7$ Mpc, and the Virgo Cluster excluded by removing
the $27$ within $10^\circ$ of the cluster centre. 
These exclusions leave a sample of $267$ spiral galaxies. 
With the correction
for galactic extinction, the resulting survey has an effective areal
coverage of $\simeq7.1$ sr (Condon 1989).

AGN contribution could potentially introduce a significant systematic
into estimates of the radio luminosity density;
indeed, Ho et al. (1997) have shown that at least $50\%$ of spiral
galaxies show some evidence for activity when a very low detection
level is used and the 
stellar absorption from the host galaxy is properly corrected. However 
the appropriate question is not whether an active nuclei are present,
but whether the nuclei contribute significantly to the low-frequency
radio fluxes. Condon, Anderson \& Helou (1991) found that removing the
Seyfert galaxies listed in the V\'{e}ron-Cetty \& V\'{e}ron (1993)
catalogue from the IRAS Bright Galaxy Sample and RSA samples, also
eliminated the outliers in the radio-far-IR correlation. 
Although such AGN segregation is somewhat arbitrary, 
this is strong circumstantial 
evidence that the remaining galaxies represent a more
homogeneous population. We can reasonably interpret the remaining
galaxies to be those
in which the radio and far-IR are dominated by star formation. 
We therefore eliminated the $36$ galaxies listed
as Seyferts in the V\'{e}ron-Cetty \& V\'{e}ron (1993) catalogue, 
differing somewhat from the approach of Condon (1989) (with a
non-negligible effect around the break of the luminosity function;
section \ref{sec:rlf}). 
The exclusion of AGN neglects any contributions from
circumnuclear star formation in these objects, and any star formation
in ellipticals is also explicitly excluded, so the luminosity
density from star formation derived in this paper should strictly
speaking be treated as a lower limit. Furthermore, it is important to
note that none of the SFR from UV estimates from the CFRS, Lyman
drop-out samples and the H$\alpha$ luminosity density of Tresse \&
Maddox (1998) (and indeed most SFR studies) 
excludes any narrow-line AGN contribution.

\begin{table}
\caption{Local $1.4$GHz luminosity function parameters: 
$\phi(L)dL=\phi_*(L/L_*)^\alpha\exp(-L/L_*)d(L/L_*)$ Mpc$^{-3}$. The
large scale structure density fluctuations are estimated to be around
$2\%$ on the scale of the survey, and are not incorporated into the
tabulated errors. We assume a Hubble constant of $H_0=50$ km s$^{-1}$
Mpc$^{-1}$.}
\begin{tabular}{lll}
Parameter & min-$\chi^2$ value & $\pm1\sigma$ range\\
          &                    &                   \\
$\phi_*$ (Mpc$^{-3})$  & $4.9\times10^{-4}$ & $(3.7-5.9)\times10^{-4}$\\
$L_*$ (W Hz$^{-1}$)   & $2.8\times10^{22}$ & $(1.9-3.3)\times10^{22}$\\
$-\alpha$  & $1.29$                        & $1.21-1.35$ \\
$L_*\phi_*\Gamma(2+\alpha)$ & & \\
(W Hz$^{-1}$ Mpc$^{-3}$) & $1.73\times10^{19}$ & $(1.35-2.08)\times10^{19}$\\
\end{tabular}
\end{table}

\section{The radio luminosity function}\label{sec:rlf}\label{sec:results}

In figure \ref{fig:rlf} we plot the $1/V_{\rm max}$ estimate of the radio
luminosity function (Schmidt 1968) for this sample. Accessible volumes
were limited on one side by the Local Group redshift limit (section
\ref{sec:method}), and on the other by the maximum redshift at which
an otherwise identical galaxy would be observable with these flux
limits. (In practice the optical limit dominates the selection.) 
As noted in Oliver, Gruppioni \& Serjeant 1998, the presence of two
flux limits (optical and radio) does not affect the $1/V_{\rm max}$
estimate of the luminosity function, as long as (a) the minimum
$z_{\rm max}$ redshift of the two flux limits is used, and (b) the
luminosity function is corrected for any parts of the (radio, optical) 
luminosity plane which are inaccessible at all redshifts. In this case 
the latter consideration is negligible.
Bins with only one galaxy have errors
corresponding to $1\sigma$ limits on the mean of the appropriate
Poisson distribution; errors are $\sqrt{\Sigma1/V^2_{\rm max}}$
otherwise. To avoid
two consecutive single-membered bins the two faintest bins were
combined to make a single bin at $10^{18\pm0.5} h^{-2}_{50}$ W Hz$^{-1}$. 
We obtained a best-fit solution with a reduced
$\chi^2$ of $0.8$, tabulated in table 1. By comparison, the original
Condon (1989) luminosity function has 
$\chi^2_\nu=5.5$ with this data set, due to the inclusion of several 
AGN in their fit.  
Errors on each parameter
were obtained treating the other parameters as fixed. 

We have already argued that the $\phi_*$ is insensitive to density
fluctuations in this sample, though it remains to be shown that $L_*$, 
$\alpha$ and the luminosity density are unaffected. In order to
demonstrate this, we compared the observed radio luminosity
distribution 
with that expected for the model luminosity function, using a fit to
the radio-optical correlation given in Condon (1987). This has the
advantage of being independent of the local density distribution. 
Using the
Kolmogorov-Smirnov statistic on the luminosity histograms to define a
likelihood statistic, we find that our min-$\chi^2$ parameters are an
excellent fit to the luminosity distribution. 


The faint-end slope is in good agreement with that obtained for the
local H$\alpha$ luminosity function by Gallego et al. (1995). 
Note that we should not necessarily expect the slope of the radio
luminosity function, which is a measure of the recent star formation
rate, to match that of {\it e.g} the B-band luminosity function, due
to the latter having contributions from less recent star formation. 
It is also important to note that our constraint on the faint-end
slope is dominated by objects with  
luminosities $(3\times10^{-3} \sim 0.3)\times L*$, and is affected
little by the exclusion of fainter bins. The data do not fit a flat
slope throughout $(3\times10^{-3} \sim 0.3)\times L*$ (though in the
faintest bins a flattening cannot be excluded).
This ``useful'' faint-end luminosity range is $\sim1$dex deeper
relative to 
$L_*$ than that of Gallego et al. 1995, and comparable to e.g. that of
Tresse \& Maddox 1998. 
The total
luminosity density of a Schechter luminosity function is given by
$\phi_* L_* \Gamma(2+\alpha)$, but the errors in these parameters are
highly correlated. We therefore explored the parameter space to find
the regions where $\Pr(\chi^2_\nu)<0.68$, and hence determined the
$\pm1\sigma$ luminosity density range, also listed in table 1. 

At these frequencies the radio luminosity is dominated by non-thermal
emission from relativistic electrons (Condon 1992) accelerated by
supernovae from massive ($M\geq5M_\odot$) stars. The lifetimes of
such stars are only $\sim10^7$ yr, so the low frequency radio emission
should be proportional to the recent star formation rate.
Following Condon (1992) and Condon \& Yin (1990), this 
can be calibrated using the Galactic non-thermal radio luminosity,
obtaining 
\begin{equation}\label{eqn:condon}
SFR(100 \geq M/M_\odot \geq 5) = \frac{L_{1.4~GHz}~[W
  Hz^{-1}]}{5.3 \times 10^{21} (\frac{\nu}{GHz})^{-\alpha}} ~M_{\odot}
  yr^{-1} 
\end{equation}
where $\nu$ is the frequency and $\alpha$ is the non--thermal radio spectral
index ($\alpha=-d\log S_\nu/d\log\nu\simeq0.8$). Assuming a Salpeter
initial mass function 
(IMF; $\psi(M)\propto M^{-2.35}$) 
we can correct to $0.1-125$ $M_\odot$ if 
we divide by $(100^{-0.35}-5^{-0.35})/(125^{-0.35}-0.1^{-0.35})=0.18$, 
and can correct from $1$ GHz to $1.4$ GHz with the factor $1.4^{-0.8}$. 
The non-thermal contribution at $1$ GHz is $\sim90\%$ of the total
luminosity (Condon 1992), so 
we therefore obtain for
the star formation rate
\begin{equation}\label{eqn:radio2sfr}
SFR(125 \geq M/M_\odot \geq 0.1) = 
  \frac{L_{1.4~GHz}~[W Hz^{-1}]}{8.2\times10^{20}} ~M_{\odot}
  yr^{-1} 
\end{equation}
where $L_{1.4~GHz}$ is the total monochromatic luminosity at that
wavelength. 
Our observed luminosity density (table 1) therefore corresponds to a
volume-averaged SFR of $(2.10\pm0.45)\times10^{-2}$ M$_\odot$
yr$^{-1}$ Mpc$^{-3}$. 
The SFR is dominated by objects around
$L_*$, peaking at $\sim0.7L_*$ corresponding to around $25 M_\odot$
yr$^{-1}$. 

We can compare this estimate with that obtained from the local
H$\alpha$ luminosity density quoted by Gallego et al. (1995). 
For our adopted IMF, Madau et al. (1998) quote a conversion of 
\begin{equation}\label{eqn:halpha2sfr}
SFR(125 \geq M/M_{\odot} \geq 0.1) =
  \frac{L_{H\alpha}~[W]}{1.41\times10^{34}} ~M_{\odot}
  yr^{-1} 
\end{equation}
for unreddened (i.e. dust-free) H$\alpha$ luminosity
densities. Gallego et 
al. dereddened their H$\alpha$ fluxes using Balmer decrements, and
their resulting luminosity density corresponds to a volume-averaged
SFR of $(8.8\pm0.1)\times10^{-3}$ M$_\odot$
yr$^{-1}$ Mpc$^{-3}$, about a factor $\times2$ lower than our radio 
estimate. We will discuss the reasons for the Balmer decrement failing 
in section \ref{sec:discussion}; 
a similar systematic offset was seen in the (albeit inhomogeneous)
samples of Cram et al. (1998), whose H$\alpha$-derived star
formation rates were a few to $10$ times lower than radio estimates,
with the more luminous systems showing the larger offset. Indeed our
luminosity function is in good agreement with that of Gallego et
al. at the faint end (figure \ref{fig:rlf}), but shows a consistently
higher number density around and above our $L_*$ where the contribution
to the luminosity density is significant. 

\section{Discussion}\label{sec:discussion}
We have found that even the de-reddened H$\alpha$ luminosity density
underestimates the local volume-averaged star formation rate by a
factor $\sim\times2$. (Strictly speaking our radio SFR is a lower
limit, as it excludes the SFR from irregulars and
ellipticals.) 
However, this does not imply that the global SFR is
dominated by extreme, highly obscured ultraluminous galaxies -- indeed
we have found the local luminosity density dominated by only $\sim25$
M$_\odot$ yr$^{-1}$ systems. Neither does it imply, for instance, that
spiral disks are predominantly opaque. It does however
require that 
a significant fraction of star formation occurs in the cores of giant
molecular clouds, which must survive destruction from
photodissociation and supernova-driven shocks long enough to obscure a
significant fraction of the O and B stars. 

At faint radio luminosities, where we have suggested the lowest
extinction, the radio--far-IR correlation becomes
non-linear. This is usually attributed to the increasing cirrus contribution
in the far-IR. This contribution can be estimated using the B
magnitudes (Devereux \& Eales 1989) but at the cost of a greatly
increased scatter in the corrected correlation. However, 
Condon, Anderson \& Helou (1991) showed that this can be avoided by
allowing the cirrus contribution to also depend on the radio
luminosity - in essence, the cirrus (of whatever temperature) is
partly heated by unobscured OB stars. This is thus in perfect accord
with our lower extinction at the faint end of the
luminosity function.

Conversely, the extinction corrections are significant at around
$L_*$. One naive interpretation of our high radio SFRs would be to
invoke a second, high extinction component to the SFR. 
However, two components are not necessary. Consider a simple 
constant-density 
model where the dust in the star forming regions is well
mixed with the $H\alpha$-emitting gas. 
Following Kroker et al. 1996 and 
Thronson et al. 1990, the observed H$\alpha$ flux $S$ of the dusty
star forming region is related to 
the ``intrinsic'' H$\alpha$ flux $S_0$ which would be observed
in the absence of dust, by
\begin{equation}
S=S_0(1-e^{-\tau})/\tau
\end{equation}
where $\tau\simeq0.7A_V$ is the H$\alpha$ optical depth to the rear of 
the star forming region. 
Assuming an extinction correction of $\stackrel{>}{_\sim}1$ magnitude
was applied to the Gallego et al. (1995) SFR (Kennicut 1983, Tresse \&
Maddox 1998), our radio SFR implies 
\begin{equation}
\frac{S}{S_0} \stackrel{>}{_\sim} \frac{0.88\times10^{-0.4}}{2.1} = 0.167
\end{equation}
and hence $A_V\stackrel{>}{_\sim}9$. (It is important to stress that
this is the extinction to the rear of the cloud, and not a ``typical'' 
extinction.) The observed H$\beta$ flux can
also be derived using $\tau({\rm H}\beta)\simeq1.45\tau({\rm
H}\alpha)$. Remarkably, the predicted Balmer decrement in this model
would imply only $A_V\simeq1.1$ in this model if one (wrongly) assumed 
a simple dust screen, {\it regardless of the
true $A_V$}, due to the low-$A_V$ regions dominating the observed
Balmer line fluxes. This is in excellent agreement with the typical
observed Balmer 
decrements in H$\alpha$ surveys (Kennicut 1983, Tresse \& Maddox
1998). The reddening corrections derived 
from optical-UV colours will also be underestimates, since the
observed spectra are also dominated by the regions with the smallest
obscuration. This effect will be strongest at high redshifts, where
the rest-frame wavelengths are the shortest. Naive
reddening corrections could therefore easily introduce
redshift-dependent biases into the comoving SFR estimates. Even 
``effective'' reddening corrections based on typical starburst
optical-UV SEDs (e.g. Heckman et al. 1998)
could miss the most highly obscured components unless such SFRs are
explicitly normalised to a more isotropic indicator (e.g. Meurer,
Heckman \& Calzetti 1998, Cram et al. 1998). 


Nevertheless, the prospects for reddening-independent constraints on
the comoving 
SFR at high-$z$ are excellent, due mainly to the fact that 
the strong far-IR--radio
correlation implies 
the far-IR can also provide unbiased SFR estimates. 
For example, optical spectroscopy of 
new surveys conducted with the Infrared Space Observatory, such as 
ELAIS (Oliver et al. 2000), ISO-HDF
(Serjeant et al. 1997, Rowan-Robinson et al. 1997) and others
(e.g. Taniguchi et al. 1997, Lemonon et al. 1998), will shortly probe
the SFR to redshifts of $z\sim1-2$. Unlike the dereddened UV and
H$\alpha$ estimates, these mid-IR and far-IR selected samples are
sensitive to star formation in even the most obscured giant molecular
clouds, with in some cases independent consistency checks available
from low-frequency sub-mJy radio follow-up observations 
(e.g. Ciliegi et al. 1999, Gruppioni et al. 1999). At higher redshifts
still, 
sub-mm sky surveys are already providing strong
constraints (e.g. Hughes et al. 1998; Barger et al. 2000). 

Finally, we note that Madau et al. (1998) could reproduce the K-band 
galaxy counts in an $\Omega=1$, $\Lambda=0$ cosmology.
Our results are therefore suggestive of a non-zero
$\Lambda$,  
as already suggested by many other recent approaches (see e.g. Peacock et
al. 2001 and refs. therein).

\section*{Acknowledgements}
This research has made use of the NASA/IPAC Extragalactic Database (NED)
which is operated by the Jet Propulsion Laboratory, California Institute of
Technology, under contract with the National Aeronautics and Space
Administration. 
We would like to thank Andreas Efstathiou, Bob Mann and Michael
Rowan-Robinson for stimulating discussions
during this work, and the anonymous referee for helpful comments.

\end{document}